\documentclass[10pt,conference]{IEEEtran}

\usepackage[english]{babel}
\usepackage[utf8]{inputenc}

\usepackage[colorinlistoftodos]{todonotes} 
\usepackage{listings}

\usepackage{cite}
\usepackage{graphicx}
\usepackage[cmex10]{amsmath}
\usepackage{algorithmic}
\usepackage{array}
\usepackage{comment}

\usepackage{url}
\usepackage{hyperref}

\newcommand{\systemname}{Metrics Dashboard}

\title{
  \systemname: A Hosted Platform for Software Quality Metrics
}

\author{
\IEEEauthorblockN{
  George K. Thiruvathukal%
  \thanks{George K. Thiruvathukal is the corresponding author. Please direct all feedback to gkt@cs.luc.edu.}%
  \IEEEauthorrefmark{1},
  Shilpika,
  Nicholas J. Hayward,
  and
  Konstantin L\"{a}ufer
}
\IEEEauthorblockA{
  Department of Computer Science \\
  Loyola University Chicago \\
  Chicago, Illinois 60611-2196, USA \\
  gkt@cs.luc.edu \\
  \IEEEauthorrefmark{1}{and Argonne National Laboratory, Lemont, Illinois 60439, USA}
}
}

\IEEEoverridecommandlockouts

\begin{document}

\pagestyle{plain}

\maketitle


\begin{abstract}
There is an emerging consensus in the scientific software community that progress in scientific research is dependent on the ``quality and accessibility of software at all levels"~\cite{wssspe}. 
This progress depends on embracing the best traditional---and emergent---practices in software engineering, especially agile practices that intersect with the more formal tradition of software engineering.
As a first step in our larger exploratory project to study in-process quality metrics for software development projects in Computational Science and Engineering (CSE), we have developed the \emph{\systemname}, a platform for producing and observing metrics by mining open-source software repositories on GitHub. Unlike GitHub and similar systems that provide \emph{individual} performance metrics (e.g. commits), the \systemname{} focuses on metrics indicative of \emph{team} progress and project health.
The \systemname{} allows the user to submit the URL of a hosted repository for batch analysis, whose results are then cached. Upon completion, the user can interactively study various metrics over time (at varying granularity), numerically and visually. The initial version of the system is up and running as a public cloud service (SaaS) and supports project size (KLOC), defect density, defect spoilage, and productivity.
While our system is by no means the first to support software metrics, we believe it may be one of the first community-focused extensible resources that can be used by any hosted project.
%
\end{abstract}




\section{Introduction}
Software engineering as practiced today is no longer about the stereotypical monolithic life cycle processes (e.g., waterfall, spiral, etc.) found in most software engineering textbooks, aimed primarily at large-scale software teams. These methods are often met with resistance by small/medium sized development teams, owing to their inherent complexity and rather limited data collection strategies that predominated the 1980s (a fervent period for emerging software engineering research) until relatively recently in the mid-2000s. As is well reported in the industry, more traditional methodology has given way to agile methods. We welcome this trend but think there is room to include many of the contributions of traditional software engineering, especially as agile projects mature and enter maintenance mode (thereby becoming much more like traditional SE projects and requiring more careful change management in particular).

The discipline and practice of software engineering includes the subdomain of software quality, which has an established theoretical foundation for software metrics. In our work, we aim to demonstrate how software processes can be effective for agile software teams by using best features/practices of various models without impeding developer productivity, especially with a growing number of cloud-based solutions for hosting projects (the most prominent being GitHub), where robust data collection is taking place and is available for anyone to study.
Our overarching motivation for focusing on software quality (and measuring it) is to support software projects in applied computing areas such as scientific computing (e.g., bioinformatics) and engineering, where there is a real need to trust the software as building blocks for larger-scale science/engineering challenges. Co-author Thiruvathukal recently co-edited a volume~\cite{carver_software_2016} that examines current practices used in scientific software development teams. In this edited volume, many scientific software development teams report the use of practices such as unit/regression testing and design methodologies. Although there were few reports on metrics, the subject is one of interest to scientific software developers, especially when it comes to performance, algorithmic correctness (related to defect density), and maintainability (related to code complexity). Co-author Thiruvathukal is leading a survey to understand the metrics needs of scientific software developers, which we expect to report separately in mid-2018. 

The challenge is to identify the most effective software engineering practices from a large suite of tools and techniques and incorporate them into existing workflows. Software development teams today actively resist using a practice if it incurs any additional workload on already short-staffed teams, especially if the solution is not integrated into existing infrastructure (e.g., GitHub, used by many open source, computational science projects alike). Accordingly, our initial \systemname{} implementation can be incorporated in any existing GitHub-based workflow without requiring  developers to install any additional software. Architecturally, the \systemname{} can support any version control system, but we intentionally focused our efforts on GitHub, where the vast majority of scientific software projects we are studying host their efforts.

Software metrics are a critical tool that provides continuous insight into  products and processes, and helps build reliable software in mission-critical  environments. 
Using software metrics, we can perform calculations that help assess the effectiveness of the underlying software or process. 

The  two  types  of  metrics relevant to our work are:

\begin{enumerate}

\item Structural metrics, which tend to focus on intrinsic code properties like code complexity.

\item In-process metrics, which focus on a higher-level view of software quality, measuring information that can provide insight into the underlying software development process.

\end{enumerate}

We understand that metrics are often used to evaluate individual developer productivity rather than overall project quality and progress. For example, a large number of commits made to a project may or may not have any impact on the software quality (yet this information is displayed prominently on sites like GitHub). Optimizing for one metric could result in unintended consequences for a project. For example, these commits could be overly complex or introduce defects. Therefore, we seek to identify metrics that will be useful to the project as a whole.

Our aim has been to develop and evaluate a \systemname{} to support Computational Science and Engineering (CSE) software development projects. To this end, we have conducted the following activities:
\begin{enumerate}

\item Assess how metrics are used, and which general classes/types of metrics will be useful in CSE projects.

\item Develop a \systemname{} that will work for teams using source code management sites like GitHub, Bitbucket, etc.

\item Assess the effectiveness of the \systemname{} in terms of project success and developer attitude towards metrics and process.

\end{enumerate}


\subsection{Novelty/contribution of this work}

The \systemname{} distinguishes itself in various ways: 
\begin{enumerate}
\item It is free/open-source software distributed under the GNU Affero General Public License (AGPL); 
\item It has an extensible architecture that makes it easy to support and study additional metrics; 
\item It provides both a human-facing web application and a RESTful web service for consumption by programmatic clients; 
\item We have implemented the \systemname{} using modern web service/application technologies and a scalable and sustainable architecture;
\item It is running as a publicly available, cloud-based software-as-a-service (SaaS) instance, and users and contributors can choose to self-host their own instance; 
\item It supports batch processing for offloading longer-running analytics jobs to dedicated resources. (We have developed an Apache Spark version of our analytics services, which is the subject of a future paper.)
\end{enumerate}
While this work is part of an effort to address sustainable practices in scientific software development, we believe it to be more broadly applicable to any interdisciplinary software development community.

\subsection{The rest of the paper}

The remainder of the paper is organized as follows. 
In Section~\ref{sec:related}, we present a summary of related work. 
In Section~\ref{sec:supported-metrics}, we discuss the initial set of \emph{in-process} metrics that we are supporting in the current (live) version of \systemname{} at \url{http://luc-metrics.herokuapp.com}. 
In Section~\ref{sec:architecture}, we describe the overall system architecture and the data visualizations it supports with respect to the various metrics. 
In Section~\ref{sec:casestudy}, we present a case study based on Google's \emph{Project Go}.
In Section~\ref{sec:conclusions}, we end with a discussion of our initial conclusions and future plans.

\section{Related Work}
\label{sec:related}


%
%

Our efforts build upon the main reference works by Fenton and Bieman~\cite{fenton_software_2014} in the area of software metrics and by Kan~\cite{kan_metrics_2002} in the more specific area of software quality metrics. As part of their comprehensive study of this area, these works stake out the software metrics design space in terms of its dimensions product metrics, process metrics, and project metrics, and define software quality metrics as a crosscutting matter. More recent works study the role of software metrics in agile processes and organizations~\cite{kupiainen_why_2014,davis_agile_2015}.

There have been various efforts to study the relationship between software metrics and development processes. 
Lind and Vairavan~\cite{lind_experimental_1989}, in the context of a real-time medical imaging system, examined the relationship among various structural software metrics and the relationship between those metrics and the development process, finding that even simple metrics such as code size correlate relatively well with development effort.
More recently, Gala et al.~\cite{gala-perez_intensive_2013}, studied the ratio of the volume of messages on project developer mailing lists and issue trackers to commits; these metrics showed some promise as indicators of project health.

\emph{Defect density} and related metrics have been receiving considerable attention. 
Gupta et al.~\cite{gupta_case_2007} conducted a longitudinal case study of two related internal projects---a reusable framework and its application---and two metrics, defect density and \emph{change density}, with defect density used as a quality metric. 
Shah, Morisio and Torchiano~\cite{shah_overview_2012} carried out a metastudy that cites 19 other papers studying 109 projects and determined that further work was needed to identify effective predictors of defect density. 
In another project involving defect density, Shah et al.~\cite{shah_impact_2012} analyzed 61 projects along two dimensions---\emph{process maturity} and \emph{process type} (hybrid vs.\ waterfall)---using defect density to measure impact of these two dimensions; they observed that process maturity has a smaller impact than process type on defect density.

Another line of research has focused on the classification of metrics, especially on the basis of mining software repositories (MSR). 
In particular, Saraiva et al.~\cite{saraiva_towards_2013} present the case for a catalog of object-oriented software maintainability metrics and made an initial attempt to classify a large number of (primarily structural) maintainability metrics on the basis of 138 primary studies. 
Similarly, Bouwers et al.~\cite{bouwers_towards_2014} argue that a detailed, formally structured catalog of metrics will enable users to make an informed decision as to which metrics to use; in addition, they see value in representing the relationships among metrics. 
Ohira et al.~\cite{ohira_dataset_2015} argue that it is important to distinguish low-impact defects from high-impact ones; to facilitate further research in this area, they used manual classification to produce a dataset of defects in several major Apache projects. 
Chaturvedi et al.~\cite{chaturvedi_tools_2013} conducted a meta-analysis of about 150 papers from all past MSR and other relevant conferences that used data mining for the experimental analysis of software engineering projects; 
they classified the data mining tools used in or developed for these projects, as well as the data mining tasks for which these tools were used, finding that about half of these data mining tasks are classification tasks and about two thirds of the tools used are common existing statistical tools.

There are few actual tools with a focus on in-process metrics. Sharma and Kaulgud~\cite{sharma_pivot:_2012} present PIVoT, a proprietary, in-house tool that supports the automated, non-invasive collection and analysis of in-process and project health metrics.
Gousios and Spinellis~\cite{gousios_platform_2009,gousios_ghtorrent:_2012} have similar goals as our effort in terms of valuing both product and process metrics and developing a working, publicly available, open-source tool or service for software engineering research. 
Their earlier tool predates the widespread availability of GitHub and focuses on open-source project repositories via conventional version control and issue tracking systems. 
Their current tool, \emph{GHTorrent}, mines half a year's worth of GitHub commits from hundreds of thousands of GitHub repositories and makes the resulting linked data available for "deep crawling" in the form of a public MongoDB database.


\section{Supported Metrics}
\label{sec:supported-metrics}
In this section, we describe the metrics supported by the \systemname{}. We begin with a general discussion of how we sample data (applicable to all of our metrics), followed by an in-depth discussion of the process-related metrics we support in the current \systemname{} implementation. 
\subsection{Longitudinal metrics and sampling}

Any process, including a software development process, occurs over time, and is therefore \emph{longitudinal} in nature. In our work, we aim to evaluate the development process through metrics that must themselves be longitudinal; i.e., they are functions of time. For example, code size may change over time whenever a committer inserts or deletes portions of the code. One way of measuring code size is lines of code (LOC), which is still used today in many projects (and even on GitHub).

While we can conceive these metrics as continuous functions of time, it is impractical to treat them as such, and one typically uses \emph{sampling} to convert them to discrete functions of time. 
The choice of sampling rate (frequency) is a practical one. If one wanted to observe, say, intra-day phenomena, one would choose a relatively high sampling rate, such as hourly or even every 15 minutes. Our research, however, focuses on longer-term phenomena, so the commonly used daily sampling rate will be sufficient. In practice, daily measurements are taken at midnight local time (00 hours).

Less frequent samples can always be obtained by \emph{downsampling} (decimation) as follows. The measurement for a metric $y(d)$ for a calendar interval $[d_0; d_1]$, where $d_i$ are calendar dates such that $d_0 \leq{} d_1$, is given as the daily average of $y$ over the time interval:
\begin{equation}
y([d_0; d_1]) = \frac{ \sum_{i = 0}^{d_1 - d_0} y(d_0 + i) }{ d_1 - d_0 + 1 }
\end{equation}
Using this definition, we can obtain measurements by week, month, or other arbitrary period.

\subsection{Metrics supported in our dashboard} 


\subsubsection{Project size}

Before we can speak about many software metrics, it is important that we discuss a common metric that is used by other metrics, including issue density (covered in the next section): the \emph{code size}, which is focused on using standard \emph{source lines of code} (LOC) counting. Our intent is not to debate the merits of different LOC counting mechanism (physical, logical, or non-comment LOC), here, because this choice itself is an extensible aspect of our architecture.

A project consists of one or more files, each with its own commit history. 
In this paper, we group longitudinal metrics with granularity of week or month, and we calculate the project size over a time range $[t; t']$ corresponding to the requested granularity. 
Accordingly, we calculate the project size by looking at the commit history of each file in the project and partitioning the commit history based on the requested granularity (a window of time).

\begin{figure*}
\centering
\includegraphics[width=0.69\textwidth]{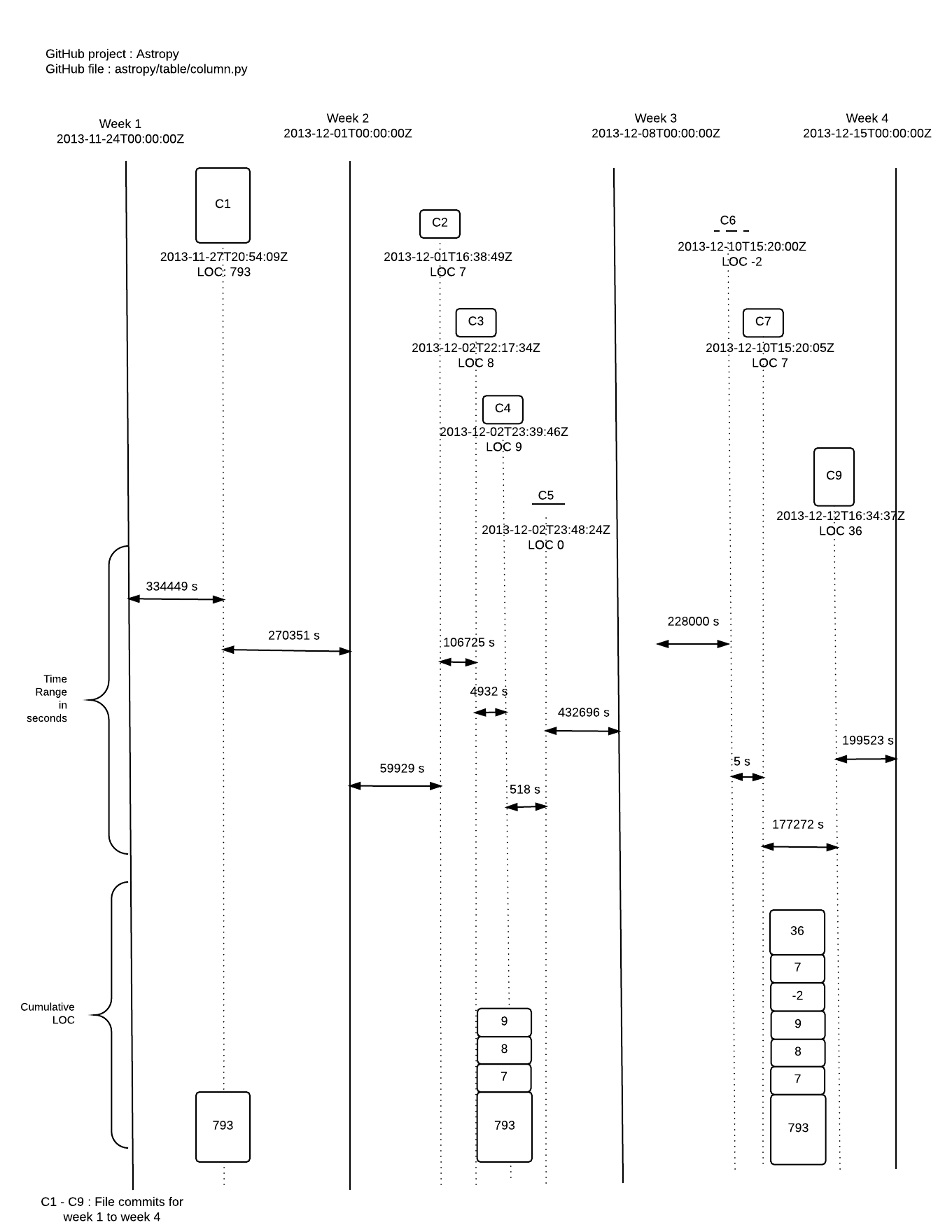}
\caption{Calculation of KLOC over the commit history of a project}
\label{fig:kloc-calc}
\end{figure*}


Figure~\ref{fig:kloc-calc} shows how to calculate \emph{file size} $s_f$ (in LOC) for the file \texttt{column.py} in a GitHub project named Astropy (chosen from projects we're actively tracking). 
We consider this file's commit history from 24 Nov 2013 to 15 Dec 2013 with weekly granularity. 
The \emph{cumulative} LOC for this file from the beginning of week one until the first commit on 27 Nov 2013 is $s_0 = 0$. Week one shows one commit for the file, resulting in the \emph{incremental} LOC value ${\Delta s}_1 = 793$ from its first commit date (27 Nov 2013) to its second commit date (1 Dec 2013).
On the second commit, we add the incremental LOC of ${\Delta s}_2 = +7$ to the cumulative LOC $s_1 = 793$ of the previous commit, resulting in $s_2 = 800$ between the second commit date to the third commit date. The process continues for subsequent commits on the file within the given time range. 
Note that the incremental LOC can contain zero or negative values; we simply add the signed LOC value for each commit to the previously calculated cumulative LOC. Owing to the longitudinal nature of our computation, we calculate the elapsed time between consecutive commits in milliseconds and later convert this to seconds for visualization of results, giving us a cumulative LOC multiplied with the time range $[t_i; t_{i+1}]$ from the current commit at $t_i$ to the next commit at $t_{i+1}$. The last size value of the file during time range $[t; t`]$ is $s_n$ at the time $t_n$ of the last commit. Finally, by dividing the combined cumulative result by the time interval corresponding to the requested granularity, we end up with a project size in LOC instead of seconds-LOC. For \emph{project size} $s_p$, we add the corresponding values for each file in the project, using same granularity window (week or month). 

In other words, we can calculate the size $\bar{s}_{f,[t; t']}$ of a file $f$ during a time range $[t; t']$ as the weighted mean size of $f$ over that time range:
\begin{eqnarray}
\frac{ 
  s_{f,t} (t_1 - t) + ( \sum_{i = 1}^{n - 1}{s_{f,t_i} (t_{i+1} - t_i)} ) + s_{f,t_n} (t' - t_n)
}{ t' - t }
\end{eqnarray}
where $n$ is the number of commits for $f$ during $[t; t']$ and $t_1, \dots, t_n$ are the times of these commits.

The size $\bar{s}_{p,[t; t']}$ of a project $p$ during a time range $[t; t']$ is then given as the sum of the sizes of all source files $f_1, \dots, f_k$ in $p$ during that time range.
\begin{eqnarray}
\bar{s}_{p,[t; t']} = \sum_{j=1}^{k}{\bar{s}_{f_j,[t; t']}}
\label{eqn:project-size}
\end{eqnarray}

The calculation of project size is crucial for the metrics calculations of issue density and team productivity covered in Sections \ref{ssec:density} and \ref{ssec:productivity}, respectively.

\subsubsection{Issue density}
\label{ssec:density}

Defined as the number of issues in a project during a specific time period of development divided by project size.
At any given time, the issue density $\rho$ of a project is given as the number of reported issues per project size $s$ in KLOC.
\begin{eqnarray}
\rho = |D| / s
\end{eqnarray}
where $D$ is the set of reported issues in the project at time $t$ and $|D|$ is its cardinality.

Given our focus on open source projects in GitHub, we actually refer to this metric as \emph{issue density}, as opposed to the more specific term \emph{defect density}, as the original research applies to any issues, not just bugs. 
GitHub provides a feature for tracking tasks, enhancements, and bugs for a project, which means that not all reported issues are actually defects. 
Having high issue density, regardless of confirmed defects, could be a sign of an unhealthy project. 
For example, if there are a large number of feature requests that are never acted upon in any way, it means that the software is lacking many features the users want. 
This would be a possible warning sign for poor customer support and, eventually, would lead to low customer satisfaction (as argued by Deming~\cite{scherkenbach_deming_2011} to be an overarching indicator of quality).



\subsubsection{Issue spoilage}




We define the \emph{age} of an issue $d$ at time $t$ as the time period from its creation time $t_d$ to $t$, and the \emph{issue spoilage} of a project as the mean age of unresolved/open issues in the project at time $t$.
\begin{eqnarray}
\alpha_d(t) & = & t - t_d
\\
\sigma(t) & = & \sum_{ d \in D }{ \alpha_d(t) } / |D|
\end{eqnarray}
where $D$ is the set of issues in the project at time $t$ and $|D|$ is its cardinality.

These metrics are useful in determining project health and team effectiveness. While we learn a great deal about issue density metrics, not attending to other non-defect issues (e.g., enhancement requests) is something developers would like to know, if only to improve customer satisfaction.



\subsubsection{Productivity}
\label{ssec:productivity}

The most commonly used model for productivity measurement expresses productivity as the ratio of \emph{process output influenced by personnel} divided by \emph{personnel effort or cost during the process}. We define productivity $P$ in the following, team-oriented way:
\begin{eqnarray}
P = \bar{s} / E
\end{eqnarray}
where project size $\bar{s}$ is defined in equation~\ref{eqn:project-size}, and team effort $E$ is calculated as the number of person hours assigned to the project during the time period of interest, $[t;t']$.

\section{The \systemname{} System}
\label{sec:architecture}

In this section, we present a detailed description of the system architecture underlying the \systemname{} system. We have architected the system with the hope of using best practices in web (services) design, especially when it comes to extensibility and supporting proper separation of concerns. We also discuss the data visualizations, which allow one to overlay various metrics to obtain a long-term perspective for any GitHub project. The web service and the visualizations support the case study in the next section.


\subsection{Architectural overview}

\begin{figure*}[!ht]
\centering
\includegraphics[width=\textwidth]{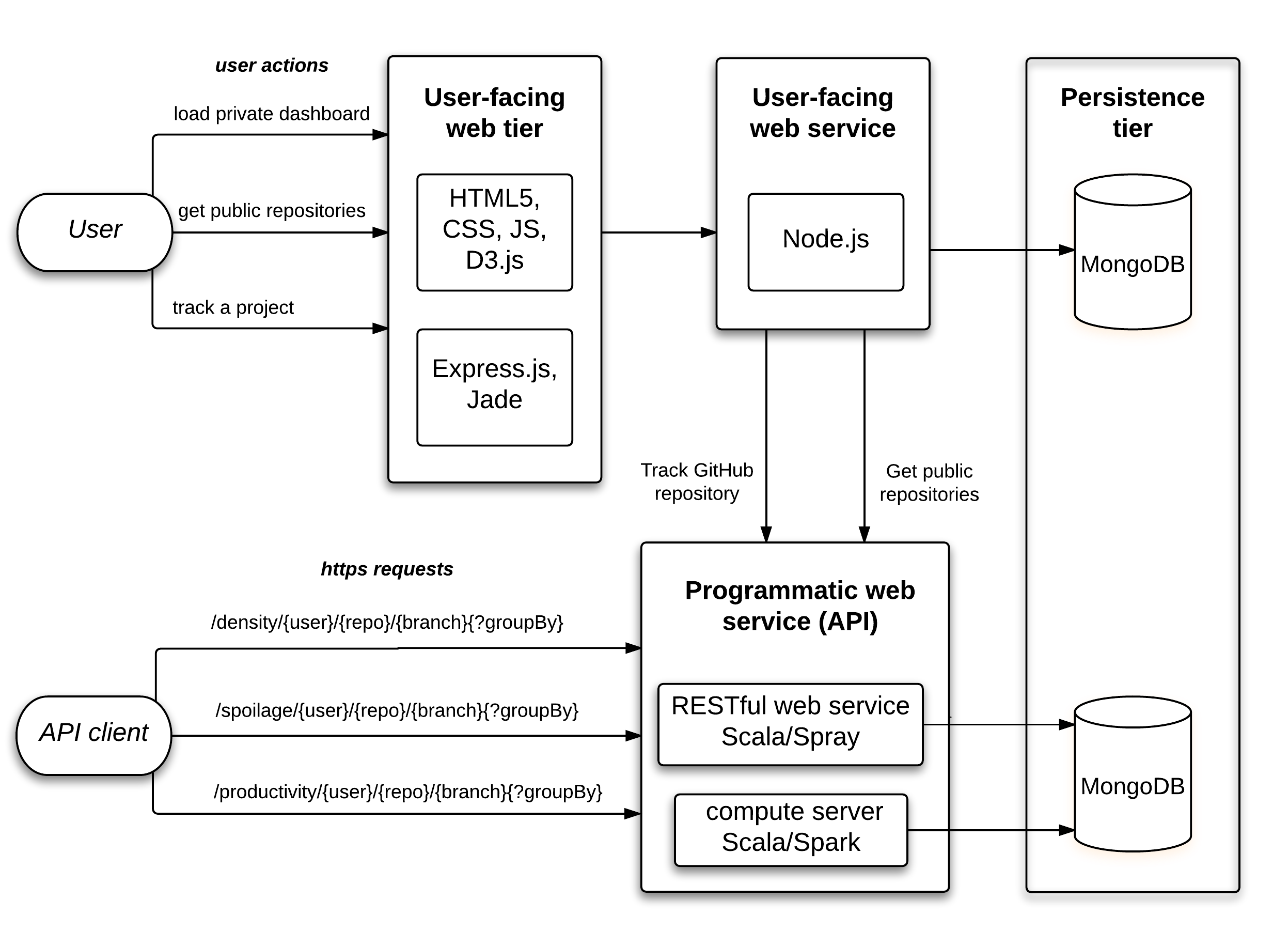}
\caption{\systemname\ system architecture}
\label{fig:architecture}
\end{figure*}

The overall system architecture is shown in Figure~\ref{fig:architecture} and comprises four general components:
\begin{itemize}
\item web tier (upper left)
\item user services tier (upper center)
\item computational services tier and RESTful API (lower left and center)
\item persistence tier (rightmost column)
\end{itemize}

Each of these tiers is responsible for different aspects of the \systemname{}. The web tier acts almost entirely as a presentation layer and also exposes user functionality (optionally) for requesting the addition of projects to the dashboard. Any project that is both open source and public can be tracked. Our current implementation only supports login with GitHub, given that the only projects we're tracking are hosted on GitHub. Future work will support other major services, such as Bitbucket and GitLab.
%
%

The web tier is separated from a layer (written in Node.js) that interfaces to the services tier, specifically when it comes to the user-specific requests to track new projects and to get the (current) results of any given repository. We provide this separation because there is user-level metadata that is only needed by the web tier, which we want to keep separate from the computational tier. 
%
%

The computational tier is where we compute and update various metrics with some degree of regularity. The analysis of some large repositories (e.g. the Linux kernel, which is several GB) simply cannot be done in real time. For these projects, we run them on dedicated cloud instances and publish the results (to the persistence layer) when they become ready. This also allows the web tier to remain responsive at all times without being burdened by long-running computations.

The persistence layer is responsible for taking the results of a computation and ensuring it is stored in the underlying database, presently MongoDB. We decided on MongoDB because our web services---and ultimately the web tier itself---all work with JSON (JavaScript Object Notation). MongoDB is also precisely designed for this sort of workflow and allows us to perform essential querying/transformation using JavaScript-based NoSQL queries. 
%
%
%
%

\subsection{Data visualizations}

A primary focus of this application is naturally the provision of a dashboard for visualizing a project's metrics data. As such, the \systemname{} has been designed for use in different contexts for a public and private user. 

\subsubsection{Public dashboard}

A public dashboard has been developed to provide data visualization of key metrics, as outlined above. This dashboard requests a given project's data from the exposed web service API, thereby consuming a project's metrics data with real-time updates. A public user is able to select an overview dashboard for key project metrics, and then refine the dashboard view for a specific metric. 

The dashboard itself is interactive, thereby permitting a user to quickly and easily filter and discriminate data based on date ranges. For example, a pie chart of project years enables quick filtering by a single year or multiple years. Likewise, each dashboard view is predicated on a frequency option for the metrics' data, including by week and month. Contextual dashboard links are offered above the visualizations, which enable quick selection of required metric and tracked project frequency. 

\subsubsection{Private dashboard}

In addition to the above options in the public dashboard, an authenticated private user is provided with a customizable private dashboard. This customization includes two primary domains, notably any required dashboard visualizations and personal projects. A private user is, therefore, able to select their preferred visualization type for their customized dashboard.

\subsubsection{Sample visualizations}

Each dashboard, both public and private, includes default visualization types, which are designed to aid in quick and easy comprehension of the selected data and frequency.

\begin{figure}[!ht]
	\centering
	\caption{Example line chart for issues open by month}
	\includegraphics[width=\columnwidth]{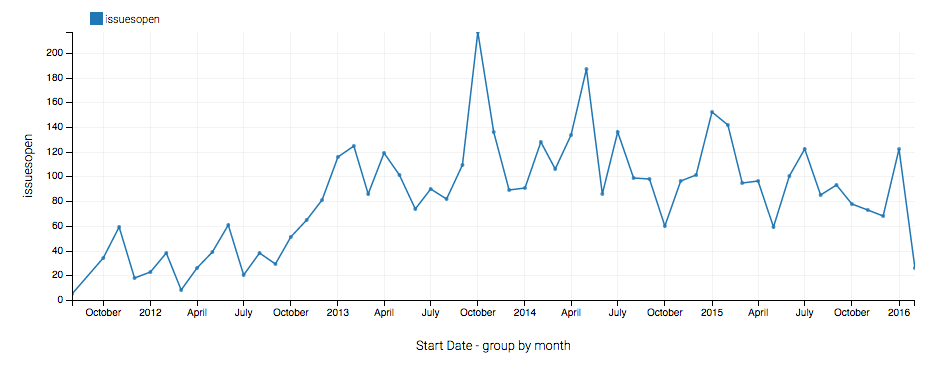}
    \label{fig:issuesopen-linechart}
\end{figure}

\begin{figure}[!ht]
	\centering
	\caption{Example comparison line chart for issues open against issue density}
	\includegraphics[width=\columnwidth]{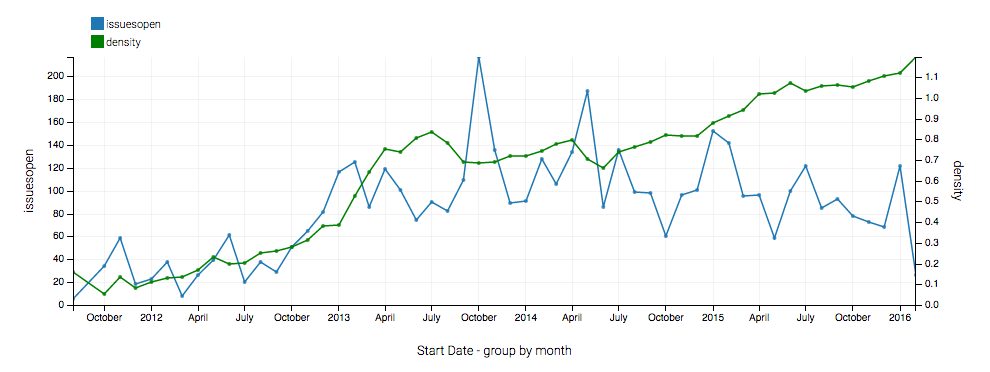}
    \label{fig:issuesopen-linechart2}
\end{figure}

\begin{figure}[!ht]
	\centering
	\caption{Example multi-comparison chart for issues open against issue density}
	\includegraphics[width=\columnwidth]{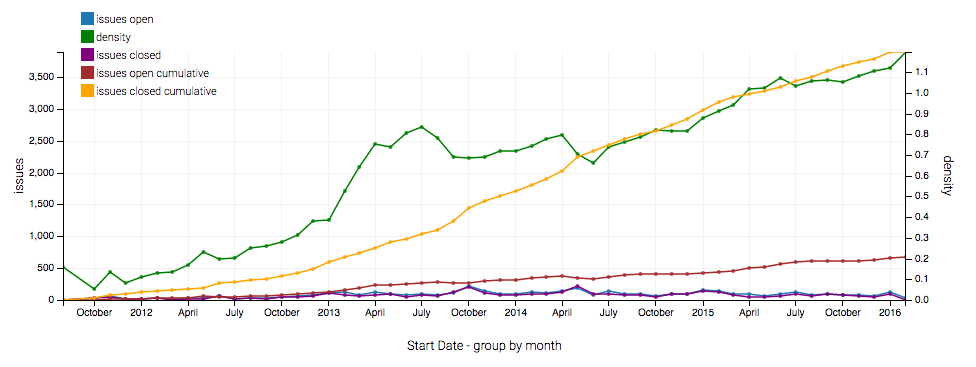}
    \label{fig:issuesopen-multichart}
\end{figure}

\begin{figure}[!ht]
	\centering
	\caption{Example pie chart for years in dataset}
	\includegraphics[width=0.8\columnwidth]{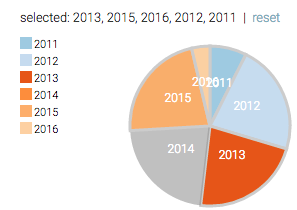}
    \label{fig:issuesopen-piechart}
\end{figure}

Current default visualization types for chosen metric and frequency include,
\begin{itemize}
\item \emph{pie chart}: presents clear division of years within a tracked project's dataset, e.g., Figure~\ref{fig:issuesopen-piechart}
\item \emph{single line chart}: initial visualization, e.g., issue density by month as in Figure~\ref{fig:issuesopen-linechart}
\item \emph{bar chart}: alternative visualization type
\item \emph{dual-comparison line chart}, e.g., issue density against KLOC as in Figure~\ref{fig:issuesopen-linechart2}
\item \emph{multi-comparison line chart}, e.g., issues open, issues closed, open and closed cumulative issues against issue density as in Figure~\ref{fig:issuesopen-multichart}
\end{itemize}

Each dashboard also features a structured table of metrics data, which includes comparative data, chosen metric data, date, and year. 

For example, we can view this combination of visualization types for ``issues open'' per month for a chosen project. We can start with our default line graph, as shown in Figure~\ref{fig:issuesopen-linechart}, then start to compare ``issues open'' against ``issue density'' in Figure~\ref{fig:issuesopen-linechart2}, and finally view a multi-comparison line graph for each ``set of open issues compared against issue density,'' as shown in Figure~\ref{fig:issuesopen-multichart}.
%
%
A user may also choose to filter the dashboard visualizations by years, as represented in a pie chart in Figure~\ref{fig:issuesopen-piechart}.

An example dashboard for the 'issues open' metric with monthly granularity is available at \url{https://luc-metrics.herokuapp.com/dash/public/month/issues/open/astropy}.

\section{Case Study}
\label{sec:casestudy}

\begin{figure}[!ht]
	\centering
	\caption{Project Go: Line chart for density against KLOC by month}
	\includegraphics[width=\columnwidth]{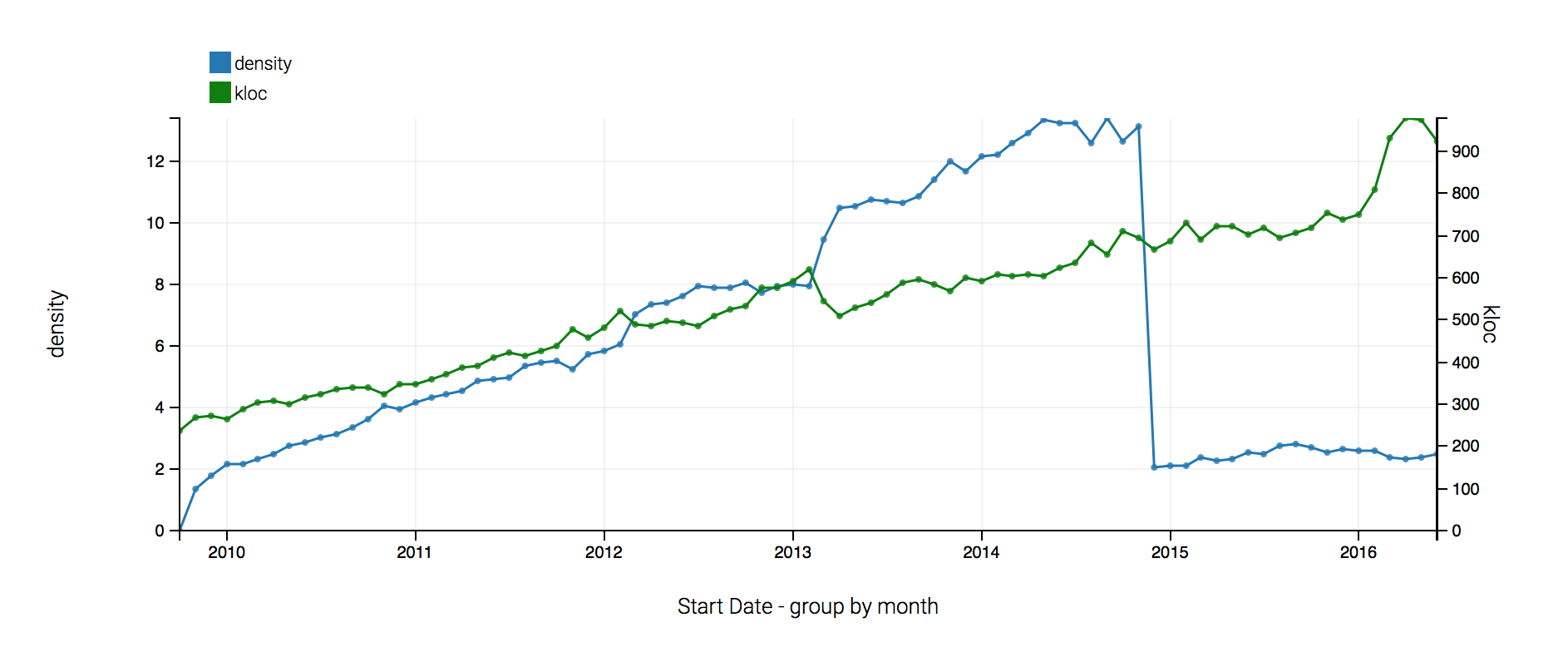}
    \label{fig:go-density-kloc-month}
\end{figure}

In this section, we present a case study based on \emph{Project Go}, an open-source programming language developed by Google and the repository is hosted in GitHub. Figure~\ref{fig:go-density-kloc-month} shows the issue density against KLOC for the project by month.

At first glance, we note a steep drop in the issue density. This drop occurs without any corresponding changes in the KLOC for the code, which suggests that a large number of issues were resolved in a short period of time. We decided to explore this further to understand what might have happened in the project and whether it is a cause for concern. Here is what we did to explore further:

\begin{itemize}
\item Navigate to the GitHub issues section for the project (\url{https://github.com/golang/go/issues?q=sort:created-asc}). Here, we notice that the first issue for the project was created in October of 2009.
\item Check the metrics dashboard service (\url{https://tirtha.loyolachicagocs.org/metrics/api/density/golang/go/master?groupBy=week}), to identify the window where the drop occurred. The results obtained from the service will be in JSON format as shown below, which contains the fields \texttt{open} or \texttt{close} and \texttt{openCumulative} and \texttt{closeCumulative}, which specify the issues opened or closed in the chosen granularity (by week) and the issues that are in the open or close state in the current window of chosen granularity (by week as well).

\begin{verbatim}
 {
  {
    "start_date": "2014-11-24T00:00:00Z",
    "end_date": "2014-12-01T23:59:59Z",
    "kloc": 651.0461298714263,
    "issues": {
      "open": 30,
      "closed": 0,
      "openCumulative": 9161,
      "closedCumulative": 0
    }
  },
  {
    "start_date": "2014-12-01T00:00:00Z",
    "end_date": "2014-12-08T23:59:59Z",
    "kloc": 653.9527515172911,
    "issues": {
      "open": 34,
      "closed": 0,
      "openCumulative": 9195,
      "closedCumulative": 0
    }
  },
  {
    "start_date": "2014-12-08T00:00:00Z",
    "end_date": "2014-12-15T23:59:59Z",
    "kloc": 639.6212584045984,
    "issues": {
      "open": 120,
      "closed": 7968,
      "openCumulative": 1347,
      "closedCumulative": 7968
    }
  }
}
\end{verbatim}

On close inspection, we notice that the date 2014-12-08T23:59:59Z is when the dip occurs, also notice that \texttt{issues closed} is 0 and \texttt{closedCumulative} is 0. 
For the next window, ending on 2014-12-15T23:59:59Z, both \texttt{closed} and \texttt{closedCumulative} are 7968.
\item Navigate to GitHub issues (\url{https://github.com/golang/go/issues}) to check if the values reported by the metrics dashboard service is correct. If we filter using the criteria \textbf{closed:\textless 2014-12-08}, we see that no issues were closed before this date (\url{https://github.com/golang/go/issues?q=closed:<2014-12-08}), even though the first issue was opened in October, 2009.
\item Change the filter to \textbf{closed:\textless 2014-12-09} (\url{https://github.com/golang/go/issues?q=closed:<2014-12-09}) and we will see that 7926 issues were closed. Change the date to \textbf{2014-12-15} (\url{https://github.com/golang/go/issues?q=closed:<2014-12-15}) and you will see 7968 issues were closed, which matches the result obtained through the \systemname{} service.
\end{itemize}

So, to summarize, the Project Go team managed to close 7926 issues in one day, which is not normally considered to be a good practice, considering the fact that the team hadn't closed any of the identified issues since October, 2009. A point to note here is that the status of issues in GitHub can be changed (even for no apparent reason). In this case, however, were are concerned only with the open and closed dates of issues. Having an issue open for five years suggests a common issue anecdotally associated with open source projects: the project relies heavily on volunteers who might not be able to keep up with incoming issues and do new development at the same time. For the above analysis, the \systemname{}'s ability to adjust granularity to week level helped to narrow down the exact date when the dip occurred.  


\begin{figure}[!ht]
	\centering
	\caption{Project Go: Line chart for density against spoilage by month}
		\includegraphics[width=\columnwidth]{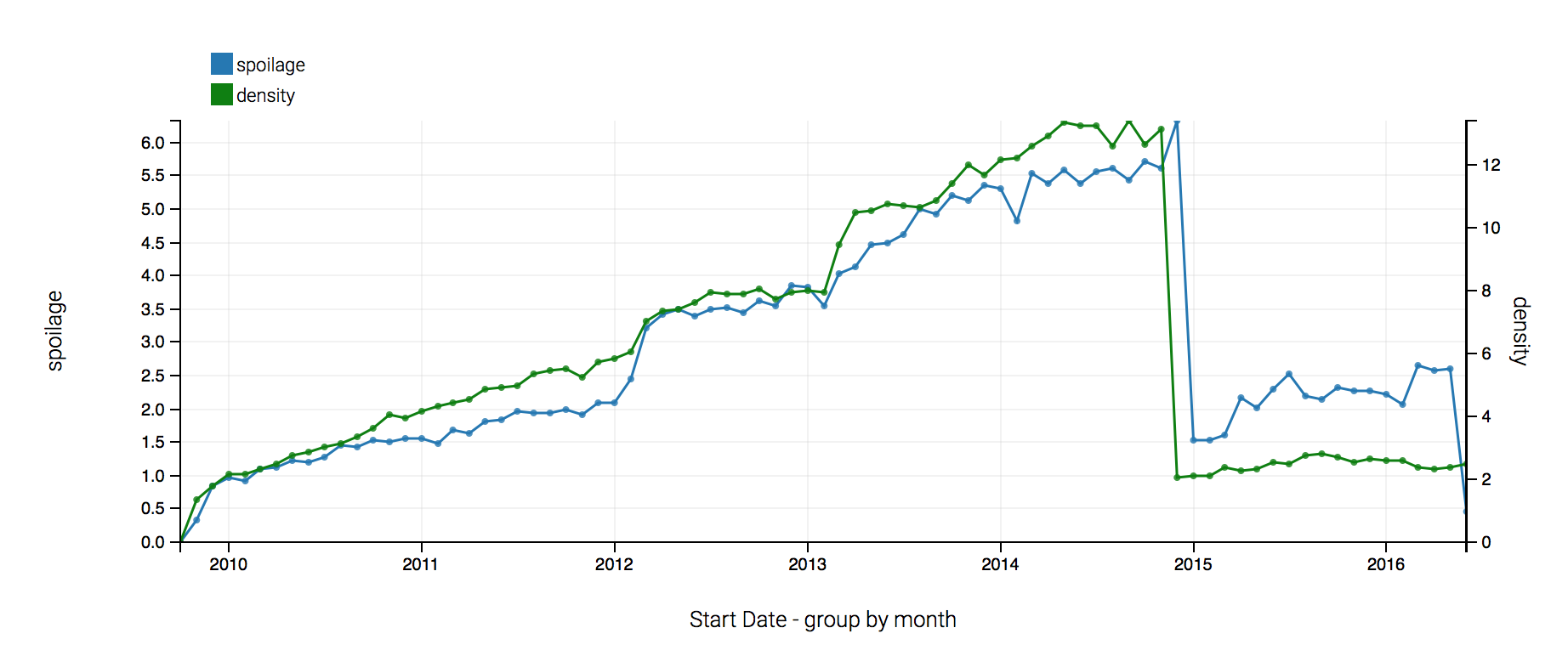}
    \label{fig:go-density-spoilage-month}
\end{figure}

Figure~\ref{fig:go-density-spoilage-month} shows the issue density against spoilage by month, as one expects the spoilage value dips at around the same window when the issue density dips. Observe that spoilage increases until the end of the year 2014 to a peak of almost 7.0, as the time to fix issues increased. After the dip the spoilage has remained constant which is a good indicator that the issues are being closed regularly and newer issues are identified and tracked. For an active project to be healthy, the spoilage should not drop too low, which could indicate that the project isn't being tested and the user community isn't actively identifying or reporting issues.

\begin{figure}[!ht]
	\centering
	\caption{Project Go: Line chart for issues grouped by week}
	\includegraphics[width=\columnwidth]{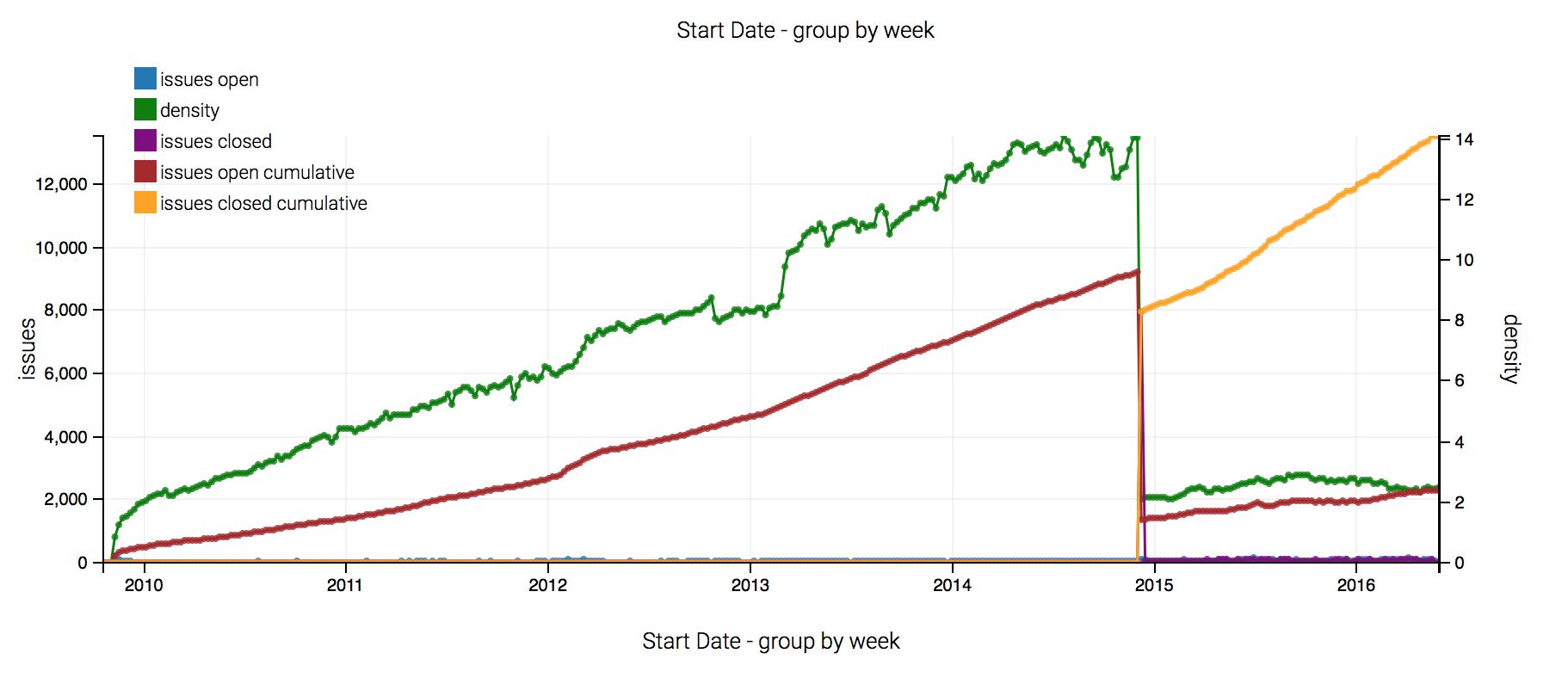}
    \label{fig:go-issues}
\end{figure}
Figure~\ref{fig:go-issues}, shows the issues for Project Go by week, as one expects the issues for the project were opened until the end of the year 2014 (cumulative open issues are shown in red). The closed cumulative issue count, shown in yellow, shows that the issues for the project were closed beginning the end of the year 2014. Since 2015, the team or users have continued to open and close issues at a fairly steady rate and no drastic changes in the values are seen. This means that the team is working to improve its work flow when it comes to resolving issues.


\section{Conclusion and Future Work}
\label{sec:conclusions}


This paper reflects our initial effort to design and develop a dashboard that can be used by project teams to assess the effectiveness of their software development process. Although at an early stage, we have already been able to use the dashboard to understand issues in existing scientific software projects that could be helpful to the project maintainers.

We are presently able to track multiple science-focused software development projects. It is not within the scope of our current paper to describe our findings about these projects. Our ultimate goal is to make it possible for teams to do their own wellness monitoring and improve their own software quality.

The current prototype at \url{http://luc-metrics.herokuapp.com} is fully-functioning and can be used by the public. Anyone with a GitHub account can log into our system and suggest new projects for us to track, which will result in them being queued for analysis and regular updating.

We're in the midst of analyzing the results from a survey of scientific software teams to assess understanding and interest in software metrics. Upon completing the analysis of this survey, we will continue work on the \systemname{} to support additional metrics and update this draft paper for conference/journal archival.


\section*{Acknowledgments}

This material is based upon work supported by the National Science Foundation under Grant ACI-1445347 and Amazon Web Services. We also acknowledge the Argonne Leadership Computing Facility for a special director's allocation of 75,000 hours of time on the Cooley/Mira supercomputers.

\bibliographystyle{IEEEtran}

\bibliography{IEEEabrv,MetricsDashboard}

\end{document}